\documentclass[11pt,fleqn,twoside]{article}
\usepackage{latexsym,amsfonts}
\usepackage[dvips]{epsfig}
\makeatletter
\newcommand{\prava}{\footnotesize\it
\begin{flushright}
\begin{minipage}{18cm}
Copyright \copyright 2009 By Author
\end{minipage}
\end{flushright}}

\newcommand{\name}[1]{\begin{flushleft}
                       \LARGE \bf #1
                       \end{flushleft}\vspace{-3mm}}

\newcommand{\Author}[1]{\begin{center}
                       \textbf{#1} \end{center}}

\newcommand{\Adress}[1]{\begin{center}
                       \it #1 \end{center}}

\newcommand{\ehkol}{Author \ name}
\newcommand{\ohkol}{Article \ name}
\renewcommand{\@evenhead}{
\hspace*{-3pt}\raisebox{-15pt}[\headheight][0pt]{\vbox{\hbox to \textwidth
{\thepage \hfil \ehkol}\vskip4pt \hrule}}}
\renewcommand{\@oddhead}{
\hspace*{-3pt}\raisebox{-15pt}[\headheight][0pt]{\vbox{\hbox to \textwidth
{\ohkol \hfil \thepage}\vskip4pt\hrule}}}
\renewcommand{\@evenfoot}{}
\renewcommand{\@oddfoot}{}

\newcommand{\be}{\begin{equation}}
\newcommand{\ee}{\end{equation}}
\newcommand{\ba}{\hspace*{-5pt}\begin{array}}
\newcommand{\ea}{\end{array}}

\makeatother

\begin{document}
\setcounter{page}{1}
\thispagestyle{empty}

\renewcommand{\ehkol}{}
\renewcommand{\ohkol}{}


\vspace{-6mm}

\renewcommand{\footnoterule}{}
{\renewcommand{\thefootnote}{}  
\footnote{\prava}}

\name{Non Homogeneous Poisson Process Model based Optimal Modular Software Testing using Fault Tolerance}\label{malek-fp} 

\Author{Amit K Awasthi and Sanjay Chaudhary}
\Adress{Pranveer Singh Institute of Technology,\\ NH-2, Kanpur-Agra Highway, Kanpur, UP, India}

\begin{abstract}
\noindent
In software development process we come across various modules. Which raise the idea of priority of the different modules of a software so that important modules are tested on preference. This approach is desirable because it is not possible to test each module regressively due to time and cost constraints. This paper discusses on some parameters, required to prioritize several modules of a software and provides measure of optimal time and cost for testing based on non homogeneous Poisson process. \\

\end{abstract}

\section{Introduction}

Whenever a software is developed a question about its reliability comes in front. We need some tool to be sure that software is working properly. That is, there is a need of software testing, to find out any faults that might exist, before releasing the product. For this purpose, software product is tested carefully but regressive testing is not feasible always, as it can be very expensive in form of cost and time both.  That's why, a modular testing is a suggestive approach so that the Testing Authority can test the software's important modules preferably and may save time and cost. 

It is impractical to test the software till all the bugs are removed, the tester should also be aware of the optimal testing time and cost required to test the modules. We also allow a bit of faults in the accepted range instead of making it 100\% error free. For this reason, this paper attempts to provide an optimal boundary values for time and cost considering the actual percentage of faults obtained in testing. A project manager should be familiar with the points where it should stop testing and go for release or rejection.

A lot of work has been done in the area of optimal software testing. McDaid and Wilson (2001) gave three plans to settle on the problem of decision - How long to test software? by introducing the optimal time measure [2]. Musa and Ackerman used the concept of reliability to make the decision [3]. Ehrlich, Prasanna, Stampfel and Wu also tried to find out the cost of a stop test decision [4]. But one of the most suitable models for the problem of determining optimal cost and time is proposed by Goel and Okumoto [5]. They gave a non homogeneous Poisson process based model to determine the optimal cost and time for software [6][7]. Praveen et al. enhanced their work by proposing a cumulative priority based elucidation to find out optimal software testing period [8].

In this paper, we consider the new idea of modular approach to test software. We suggest here to assign a weight on each modules depending on various parameters. Hierarchies of the modules also plays an imporatant role in decision as preceder module will always affact their dependent modules. We enhanced previous ideas by adding this hierachical module concept.

The next section briefly explains background and related work. Section 3 provides the module prioritization schema based on various factors and our approach to test the software to determine that the software is OK for release or not. Section 4 brings an example where this approach is applied. Last section concludes finally. 

\section{Background and Related Work}
\subsection{Non homogeneous poisson process}
A Poisson process is one of the most significant random processes in probability theory. It is widely used to model random points in time and space such as the times of radioactive emissions, the arrival times of customers at a service center and the positions of flaws in a piece of material. Several important probability distributions arise naturally from the Poisson process. The Poisson process is a collection of random variables where $N(t)$ is the number of events that have occurred up to time $t$ (starting from time 0) [8]. The number of events between time $a$ and time $b$ is given as $N(b).N(a)$ and has a Poisson distribution. A Non-Homogeneous process is a process with rate parameter $\lambda(t)$ such that the rate parameter of the process is a function of time e.g. the arrival rate of vehicles in a traffic light signal.
\subsection{Related work by Goel and Okumoto} 
Faults present in the system causes software failure at random times. Let $N(t)$ (where $t > 0$) be the cumulative number of failures at time $t$ (either CPU time or calendar time). According to Goel and Okumoto [5], Let $m(t)$ be the expected number of faults detected by time $t$ can be shown as \ref{eq1}:
\begin{equation}\label{eq1}
m(t) = a(1-e^{-bt})
\end{equation}
where, $m(\infty) = a$  so that a represents the expected number of software failures to be eventually encountered and b is the detection rate for an individual fault.

According to Goel and Okumoto, the operational performance of a system is to a large extent dependent on testing time. Longer testing phase leads to enhanced performance. Also, cost of fixing a default during operation is generally much more than during testing. However, the time spent in testing delays the product release, which leads to additional costs. The objective is to determine optimal release time to minimize cost by reducing testing time. Goel and Okumoto gave the parameters $c_1, c_2, c_3, t$ and $T$ which are as follows:

$c_1$ = cost of fixing a fault during testing

$c_2$ = cost of fixing a fault during operation $(c_2>c_1)$

$c_3$ = cost of testing per unit time

$t$ = software life cycle length

$T$ = software release time (same as testing time)

Since $m(t)$ represents the expected number of faults during $(0,t)$ the expected costs of fixing faults during the testing and operational phases are $c_1m(T)$ and $c_2(m(t)-m(T))$ respectively. Further, the testing cost during a time period $T$ is $c_3(T)$. If there is a cost associated with delay in meeting a delivery plan, such a cost could be included in $c_3$. Combining the above costs, the total expected cost is given by (2).
\begin{equation}
C(T) = c_1 m(T) + c_2[m(t) - m(T)] + c_3(T) 	
\end{equation}
This policy minimizes the average cost and depends on the ratio of $a*b$ and 
\begin{equation}
	C_r = c_3 / (c_2-c_1)
\end{equation}
Two cases arise, $ab > C_r$ and $ab \leq C_r$

Case I : If $ab > C_r$, the optimal policy is to take
\begin{equation}
	T^* = min (T_0, t)
\end{equation}
where $T_0 = 1/b ln(ab / C_r)$

Case II : If $ab <= C_r$, then $T = 0$.
If the cost of testing or cost of delay in release are very high, the solution favors no testing at all i.e. $T^* = 0$. 

On the other hand, if the cost of fixing a fault after release is very high as compared to the usefulness of the system, the solution will tend to favor not using the system i.e. $T^* = t$.

\subsection{Related work by Praveen et al.}
This paper suggests prioritizing the software modules into 5 categories namely \textsl{very high, high, medium, low and very low}. Then they calculate optimal cost and time similar to Goal and Okumoto work. To find out maximum allowable cost and time stringency concept is used here. Stringency is the maximum allowable deviation from the optimum which is decided by the organization.

Then they advise to start testing the software to calculate the actual time and actual cost for each priority category. The deviation from optimal testing time and optimal cost can be calculated from (5) and (6).
\begin{equation}
\alpha = {(T_a - T^*) \over T^*}	
\end{equation}
Where,

$\alpha$ = deviation from optimal time

$T_a$ = actual testing time

$T^*$ = optimal testing time calculated from (4), and 
\begin{equation}
\beta = {(C_a - C_0) \over C_0}	
\end{equation}
Where,

$\beta$ = deviation from optimal cost

$C_a$ = actual testing cost

$C_0$ = optimal testing cost calculated from (2)

Limiting factor $\delta$ is given by (7)
\begin{equation}
	\delta = \alpha + \beta
\end{equation}
Afterwards they cumulatively calculate the limiting factor ä to determine whether further software testing is required.
\subsection{Related work by Ohba} 
The above discussed models view the software as single unit, regardless of the structural or functional relationship among software subsystems (modules). Based on the concept of redundancy, recovery block techinique \cite{HEC79} and N-version program techinique \cite{AVI85} s-independently produce multiple versions of the software to perform the same function.

Most software reliability models assume s-independence of faults. However, Ohba \cite{OHB94} argues that faults are s-dependent because of the logical or functional dependency within a program. Ohba observed an S-shaped software reliability growth curve, as opposed to the exponential growth curve for the s-independence models. The model is characterized by:
\begin{equation}
m ( t ) = n. [ 1 - (1 + \phi.t ).exp(-\phi.t)]
\end{equation}
Unlike most software reliability models that use execution time, the S-shaped model is generally observed when calendar time is used.
\subsection{Musa-Okumoto}
Musa \& Okumoto \cite{Musa87} proposed a logarithmic Poisson execution-time model where the observed number of failures
by time $t$ is NHPP. This model adds a decay parameter, and is characterized by:
\begin{equation}
m( t ) = (1/ \theta).\log(\lambda.\theta.t+1)
\end{equation}

\section{Proposed Approach}
\subsection{Components Priority}
To ensure that the component prioritization is uniform and effective, it is imperative to introduce a schema [13]. The following parameters may  be helpful to decide the priority of the components.

\textbf{Production Time} – This is the amount of work carried out by an employee on the project. This parameter keeps the track of total person hours for a module. Module priority will increase as Production time increases.

\textbf{Decision density} – 
High complexity may result in bad understandability and more errors. Complex procedures also need more time to develop and test. Therefore, excessive complexity should be avoided. Too complex procedures should be simplified by rewriting or splitting into several procedures. Complexity is often positively correlated to code size. A big program or function is likely to be complex as well. These are not equal, however. A procedure with relatively few lines of code might be far more complex than a long one. We recommend the combined use of lines of code and complexity metrics to detect complex code. The total cyclomatic complexity for a module is calculated as follows. 
\begin{equation}
TCC = Sum (CC) - Count (CC) + 1	
\end{equation}

Cyclomatic complexity is usually higher in longer procedures. How much decision is there actually, compared to lines of code? This is where you need decision density (also called cyclomatic density). 
\begin{equation}
DD = CC / LLOC
\end{equation}
 
where LLOC id logical lines of codes. This parameter shows the average decision density of the code lines within the modules.

\textbf{Programming Path} – This parameter suggest that what environment for coding is used. Costs associated with technology required for the component. What are the importance of current technology for this component. How much experts are available for such technologies.

\textbf{Size of Components} – How much code had done?

\textbf{Skill of fault reporters/resolvers} – Source of origin of fault suggested is how much reliable. Errors are reported technically or just by inexperience of user. Actually in our model, we consider that faults are collected using some bug tracking system which is open to customer too.

\textbf{Weight priority }– This includes the ranking given by developers, managers and customer based on the requirements and previous experiences. It also includes risk factors. 

\textbf{Code reusability} – If an earlier source code can be used in the current work with little or no modifications then we call it code reusability. This lessens the requirements of testing the code again as it has already been tested earlier.

\textbf{Coupling} – It is the measure of connectedness of one module to another. It is given as-
\begin{equation}
C = 1 -– \left(k \over {(d_i + ac_i + d_o + bc_o + g_d + cg_c + w + r)}\right) 
\end{equation}

Where $C$ = Coupling

$d_i$ = number of input data parameters

$c_i$ = number of input control parameters

$d_o$ = number of output data parameters

$c_o$ = number of output control parameters

$g_d$ = number of global variables used as data

$g_c$ = number of global variables used as control

$w$ = number of modules called (fan-out)

$r$ = number of modules calling the module under consideration (fan-in)

the values of $k$ and $a, b$ and $c$ may be adjusted as more experimental verification occurs [11].

\textbf{Layout appropriateness}
For a specific layout (i.e., a specific GUI design), cost can be assigned to each
sequence of actions according to the following relationship:
\begin{equation}
	\textrm{cost} = \Sigma [\textrm{frequency of transition}(k) \times \textrm{cost of transition}(k)]
\end{equation}
where $k$ is a specific transition from one layout entity to the next as a specific task is
accomplished. Layout appropriateness is defined as
\begin{equation}
	LA = 100 \times [(\textrm{cost of LA} - \textrm{optimal layout})/(\textrm{cost of proposed layout})]
\end{equation}
where $LA = 100$ for an optimal layout.

\textbf{Maintenance}
$M_T$ = the number of modules in the current release
$F_c$ = the number of modules in the current release that have been changed
$F_a$ = the number of modules in the current release that have been added
$F_d$ = the number of modules from the preceding release that were deleted in the current release

The software maturity index is computed in the following manner:
\begin{equation}
	SMI = [M_T - (F_a + F_c + F_d)]/M_T
\end{equation}
As $SMI$ approaches 1.0, the product begins to stabilize. SMI may also be used as parameter for planning software maintenance activities.

The parameters are not limited as above. Some other parameters may also be used. Even fuzzy parametes may also included.

\subsection{Weight Parameter for Each Component}
In our system these parameters are based on neural networks. Assume that $w_{1,ij},~(i=1, 2, 3, ..., p; j=1, 2, 3, ..., q;)$ are the weight between $i$-th unit on sensory layer and $j$-th unit on association layer. And, $w_{2,jk},~(j=1, 2, 3, ..., q; k=1, 2, 3, ..., r;)$ are the weight between $j$-th unit on association layer and $k$-th unit on response layer. $x_i$ represent the normalized input variables to the $i$-th unit on sensory layer and $y_k$ represent the output values. We apply normalized values of fault level, fault reporter, etc to input values $x_i$. Cosider the logistic activation function, sigmod function
\begin{equation}
	f(x) = {1 \over {1+e^{-\theta x}}}
\end{equation}
Then the input-out rules of each unit on each layer are 
\begin{equation}
h_j = f(\sum_{i=1}^{p}{w_{1,ij}x_i})	
\end{equation}
\begin{equation}
y_k = f(\sum_{j=1}^{q}{w_{2,jk}h_{ji}})	
\end{equation}
We apply the multi-layered neural networks by propagation in order to learn the interaction among software components \cite{Kar90}. Now as the error in $y_k$ may be given as 
\begin{equation}
\epsilon_k = \frac{1}{2} \sum_{k=1}^{r}{(y_k - d_k)^2})	
\end{equation}
where $d_k$ are the target input values for the output values. We consider the estimation and prediction model so that the property of interation among software components accumulates on the connection weight of neural networks. Finally, we may obtain the total weight parameter $p_k$ which  represents the level of importance for each component 
\begin{equation}
p_k = \frac{y_k}{\sum_{k=1}^{r}{y_k}}	
\end{equation}

\subsection{Our Extension to Goel and Okumoto Scheme}

In Goel-Okumoto method, $m(t)$ represents the faults during $(0,t)$, the expected costs of fixing faults during the testing and operational phases are $c_1m(T)$ and $c_2(m(t)-m(T))$ respectively. Further, the testing cost during a time period $T$ is $c_3(T)$. If there is a cost associated with delay in meeting a delivery plan, such a cost could be included in $c_3$. 

Here we assume that software developement is in muti-version environemt. During the developement phase of current version some, fault appears in previous version. It is clear that cost to repair that fault goes to previous version's cost, which we could not include here. But fault appearing in previous version is nearly equivalent to finding fault is current version. The cost for this could not be same as $c_1$. We assume this newly associated cost as $c_4$. Now if 
$n(t)$ represents the faults in previous version during $(0,t)$, the expected costs of fixing faults during the testing and operational phases is $c_4 n(T)$. Thus, total expected cost is now
\begin{equation}
C(T) = c_1 m(T) + c_2[m(t) - m(T) - n(T)] + c_3(T) + c_4n(T) 	
\end{equation}
\subsection{Component Importance basis Testing}
Now, we decide level of priority on the basis of parameter $p_k$. In order to resolve tie cases manual decision may be prefered. If some dependent module should be given much more prefernce if its parent module is not tested. After prioritzing the modules, try to find optimum cost and time parameters in very similar way to Goel's Model.

Let $T$ and $C$ be the total time and cost available to release the software. Our aim is to the test all the modules within $T$ and $C$. But if we are not able to do this then at least the components with very high priority must be tested. We set the fault tolerance = 0 for the first time testing of all the components of a particular category (e.g. Very High) and find out actual time and cost for testing.

If optimal cost and time parameters $C^*$, $T^*$ are determined, then we can compute a expected cost as limiting factor $\delta = f(T,T^*,C,C^*)$. i.e.
\begin{equation}
\delta = p\frac{(C-C^*)}{C^*} +  (1-p)\frac{(T-T^*)}{T^*} 	
\end{equation}
\noindent where $p$ is odds in in favour of cost.

\end{document}